\documentclass{IEEEtran}
\usepackage[table]{xcolor}
\usepackage{cite}
\usepackage{graphicx}
\usepackage{lipsum}
\usepackage{stfloats}
\usepackage{amssymb}
\usepackage{pifont}
\usepackage{bbding}
\usepackage{booktabs}
\usepackage{ctable}
\usepackage{threeparttable}
\usepackage{footmisc}
\usepackage{framed} 
\usepackage{array}
\usepackage{multirow}
\usepackage{longtable}
\usepackage{rotating}
\usepackage{fancyhdr}
\usepackage{lastpage}
\usepackage{pifont}
\usepackage{layout}
\usepackage{wrapfig}
\usepackage{xcolor}
\usepackage{tabularx}
\usepackage{tabu}
\usepackage{enumitem}
\usepackage{url}

\graphicspath{{Figures/}}

\usepackage{array}

\DeclareGraphicsExtensions{.png, .eps}

\begin{document}
	\title{Knowledge-Assisted Privacy Preserving in Semantic Communication}
	\author{
	Xuesong Liu, Yao Sun,  Runze Cheng, Le Xia, Hanaa Abumarshoud, Lei Zhang, and Muhammad Ali Imran
	\thanks{
	
Xuesong Liu, Yao Sun (\textit{corresponding author}), Runze Cheng, Le Xia, Hanaa Abumarshoud, Lei Zhang, and Muhammad Ali Imran are with University of Glasgow, United Kingdom;
	
	}}	
	\maketitle
 
%\begin{table*}[ht]
   % \centering
    %\caption{KB Security Evaluation}
 %\begin{tabular}{|c|c|c|c|c|}
 %% \hline
  %Format & Structured Storage &  Secure Sharing & Semantic Inference  & Privacy Delivery\\
 %%% \hline
 %%%% Pre-training Datasets &\ding{55}& \ding{55} & \ding{55} &\ding{55}\\
 %%%%% \hline
 %%%%%% Semantic Web &\checkmark& \ding{55} & \checkmark &\ding{55}\\
  %%%%%%%\hline
%%%%%%%%% Knowledge graph &\checkmark& \checkmark & \checkmark&\checkmark\\
  %%%%%%%% \hline
%%%%%%%%%% \end{tabular}
%\end{table*}

\begin{abstract}
Semantic communication (SC) offers promising advancements in data transmission efficiency and reliability by focusing on delivering true meaning rather than solely binary bits of messages. However, privacy concerns in SC might become outstanding. Eavesdroppers equipped with advanced semantic coding models and extensive knowledge could be capable of correctly decoding and reasoning sensitive semantics from just a few stolen bits. To this end, this article explores utilizing knowledge to enhance data privacy in SC networks. Specifically, we first identify the potential attacks in SC based on the analysis of knowledge. Then, we propose a knowledge-assisted privacy preserving SC framework, which consists of a data transmission layer for precisely encoding and decoding source messages, and a knowledge management layer responsible for injecting appropriate knowledge into the transmission pair. Moreover, we elaborate on the transceiver design in the proposed SC framework to explain how knowledge should be utilized properly. Finally, some challenges of the proposed SC framework are discussed to expedite the practical implementation.

 %Semantic communication (SC) improves data transmission reliability and efficiency by focusing on accurately delivering true meaning implied in source messages. However, knowledge bases may contain personal data and sensitive information, which can lead to privacy breaches if they are inappropriately accessed or shared, thus laying down hidden dangers. We first investigate the categories of knowledge in semantic communication, and analyze potential eavesdropping attacks. To avoid these potential threats, we propose 
 
 %we design a knowledge-assisted privacy preserving semantic communication framework, which provides a new way by utilizing knowledge to protect privacy. Additionally, we propose specific processes for building private knowledge. Furthermore, a novel coding-decoding model is designed, analysis , some challenging problems and applicable solutions are analyzed.
\end{abstract}
	
\section{Introduction}
Semantic communication (SC) has been recognized as a transformative paradigm that enhances transmission efficiency and reliability by shifting the focus to the delivery of the meaning of source messages \cite{sun}. By deploying intelligent semantic coding models at both transmitters and receivers, fewer bits are required to compress source messages at a semantic level. Additionally, accurate semantics can be recovered from the received distorted bits, even under harsh channel conditions.  

Knowledge, the pillar of SC, is capable of underpinning semantics representation, extraction, and recovery \cite{r16}. Basically, knowledge is a data repository containing structured information and semantic relations, providing strong context information and reasoning ability for transmitters and receivers \cite{zhao2024enhancing}. More specifically, knowledge could assist transmitters in compressing source messages into semantics for encoding with low redundancy. At the receiver side, reasoning ability empowered by knowledge is needed to recover the true semantics from the distorted bits. Knowledge usually contains specific domain information and semantic rules. For instance, in the field of medical science, the knowledge may contain information related to diseases, symptoms, treatments, etc. Technically, the indirect relationship between diseases and treatments can be inferred via the contextual information in the knowledge, and their semantic rules are analyzed. Moreover, knowledge should be periodically updated in SC systems to keep its timeliness. 

Although the prospects of SC have been widely recognized, privacy issues exist as a vital barrier to unlocking its full potential. Eavesdropping attacks, referring to the unauthorized interception and monitoring on data transmission, bring significant challenges for privacy leakage\cite{r11}, especially in SC systems. Unlike conventional eavesdropping attacks, eavesdroppers equipped with a capable semantic coding model may be able to reconstruct semantics from only a few stolen bits\cite{r12}. Furthermore, if attackers hold partial or even full knowledge of the transceiver, the threat becomes more severe. In such cases, eavesdroppers can leverage their knowledge in conjunction with capable semantic coding models to infer additional semantics from the stolen bits, such as hidden information and future messages between the transmission pair. To resit to serious privacy leakage issues, privacy-preserving SC designs are urgently needed.

To avoid privacy leakage under the aforementioned attacks, the major challenge is to prevent attackers from restructuring source information with eavesdropped semantics. Therefore, privacy-preserving design should be placed on the two core components of SC, i.e., encoding-decoding model and knowledge. Although some researchers have delved into the privacy-preserving SC design by securing encoding-decoding models, only a few contributions highlighted the potential of preventing privacy leakage by means of tailored knowledge design and utilization. Knowledge is not only a simple database that enables the encoder-decoder to compress and restructure source messages, it also has great potential to assist in a dynamic codebook construction for a specific SC transceiver. Since the context-specific semantic relationship and semantic entity of the knowledge can be periodically updated, it is difficult for attackers to accurately synchronize the whole knowledge for restructuring the source message. In addition, knowledge could be utilized in a dedicated and pre-agreed manner for the transceiver, misleading eavesdroppers into exploiting the knowledge incorrectly and reconstructing inaccurate semantics.

In this article, we design a knowledge-assisted privacy preserving SC framework to resist eavesdropping attacks. The main contributions are summarized as follows:

\begin{itemize}
\item The categories of knowledge in SC are analyzed, followed by a discussion of three typical eavesdropping attacks in SC that can lead to privacy leakage.
\item We propose a knowledge-assisted privacy preserving SC framework by proposing novel designs for a reliable data transmission layer and a knowledge management layer.
\item We illustrate the workflow of encoders and decoders in the proposed privacy preserving SC framework, and demonstrate the ability of resisting different eavesdropping attacks. 
\item Finally, to expedite the implementation of the proposed framework, we discuss some potential challenges that may arise in practice.
\end{itemize}
\begin{figure*}[ht] % htbp表示将图片放置于当前位置、上方、下方或两侧（如果有足够空间）
    \centering
    \includegraphics[width=\textwidth]{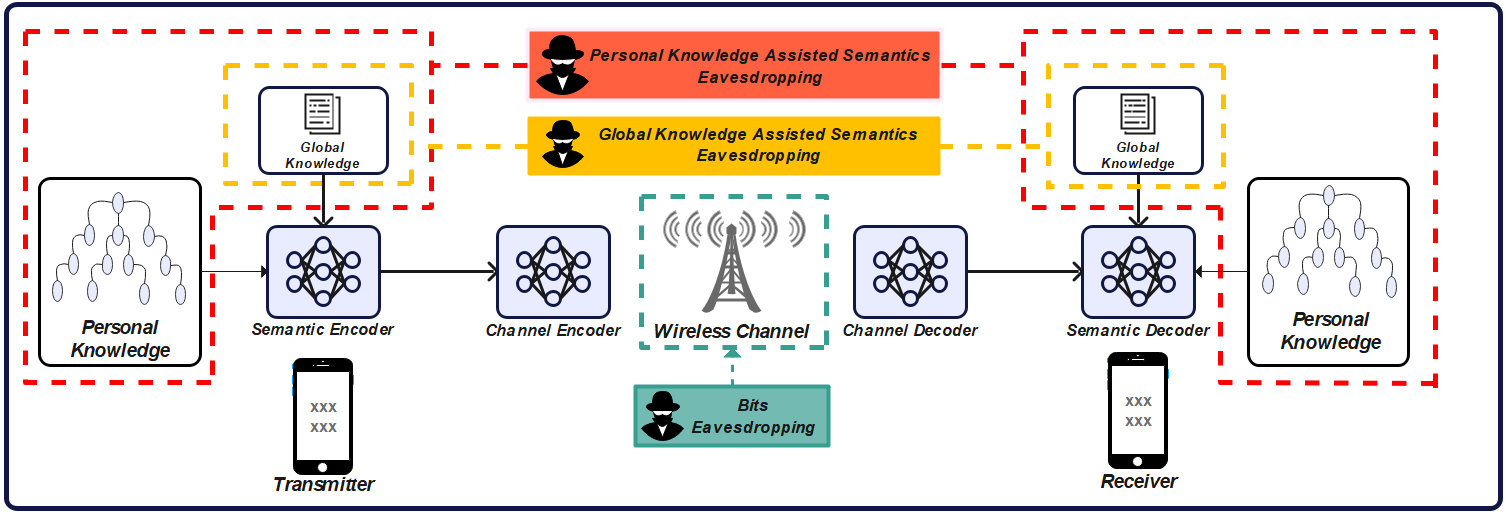} % your_image.jpg为要插入的图片路径
    \caption{Three potential attacks against knowledge in the semantic communication system.} % 添加标题
    \label{fig:1} % 设定引用标签
\end{figure*}

\section{Potential Privacy Threats in SC}
Before addressing privacy issues in SC, let us first overview the knowledge categories in SC. Following that, we define three typical eavesdropping attacks, as shown in Fig.\ref{fig:1}.

\subsection{Categories of Knowledge in SC}
As discussed before, knowledge plays a key role in SC for semantics extraction and recovery. Generally, there are two categories of knowledge in SC.

\textbf{Global knowledge} refers to public information that can be accessed by everyone, such as publicly-available knowledge from the Internet. Therefore, in SC networks, it is commonly assumed that transmitters and receivers hold the same global knowledge. By injecting global knowledge into the semantic encoders and decoders, enables two communication ends to hold the same understanding of public information.

\textbf{Personal knowledge} consists of some private and sensitive content of each individual user, such as user behavior, preferences, and the knowledge gained from individual practice. Transmitters and receivers hold their own personal knowledge. At the transmitter end, the semantic encoder exploits personal knowledge to filter out irrelevant information, reducing redundancy during encoding and further compressing the source messages. At the receiver end, the semantic decoder recovers the meaning of source messages from the received bits with the assistance of the receiver's personal knowledge. Moreover, if the recovered messages are ambiguous, personal knowledge can be supplemented to help receivers understand the semantics precisely.

%\textbf{Updated and fused knowledge} is the process of iterating and aggregating based on the knowledge of the transmitter and receiver. Updated knowledge involves the maintenance and synchronization of public and personal information for both parties, supporting the accuracy of semantic encoding and decoding, ensuring mutual understanding of the latest semantics. The purpose of fused knowledge is to establish a consensus between the transmitter and receiver regarding all information. 

%\textbf{Private knowledge} is based on fused knowledge to refine the transmitter and receiver consensus on privacy. For example, a message ``It's raining in London today'' allows attackers to infer from the global KB that the transmitter may be living in London. However, the receiver may not reside in London, and by default this message also originates from the global KB, thus the privacy definition needs to be consistent between transmitter and receiver.

\subsection{Three Eavesdropping Attacks in SC}
%Compared to traditional communication, attacker in SC need to steal deep information meanings, reflected in the semantic encoder for knowledge injection. As shown in Fig.1, attacker carrying powerful decoding capabilities (with channel decoder and semantic decoder) can decipher semantic information based on knowledge. Due to the different sets of knowledge possessed by the attackers, 
Based on the discussions on knowledge, we identify the following three types of eavesdropping attacks in SC.
\subsubsection{Bits Eavesdropping}
This kind of attack refers to stealing source messages by purely eavesdropping on wireless signals over a wireless channel. In this way, attackers can only intercept bits but may not decode accurate semantics from these bits due to the lack of a matched semantic coding model. Therefore, attackers can detect that there are some information exchanges between the two ends, but they cannot understand the information itself. Note that the knowledge attackers hold does not matter in this case due to the lack of a qualified semantic coding model.
\subsubsection{Global Knowledge Assisted Semantics Eavesdropping} Attackers in this case are equipped with a matched semantic model along with global knowledge, enabling them to eavesdrop on some public semantics. Attackers first intercept bits over a wireless channel, and then decode semantics by using their own semantic coding model, injected with global knowledge. In this way, attackers can recover the public semantics of source messages to some extent, leading to privacy leakage issues at the semantic level. Moreover, by analyzing all eavesdropped bits, attackers discover that limited bits can be decoded into public information, thus determining that the undecoded bits may contain other semantic information.
%This refers to that attackers have the semantic coding model and global knowledge, and global knowledge is injected into the semantic coding model to decode the semantics. In this case, attackers understand the same public information as  transmitters and receivers. During frequent SC, attackers use sniffing tools on the physical channel to intercept source messages and combine them with global knowledge to infer transmitter privacy. On this basis, the attacker injects malicious data into the global knowledge so that the transmitter and receiver cannot understand correct semantics, forcing the transmitter to send messages to the receiver frequently, thereby leading to more private information being snooped. 

%Transmitter knowledge attacks refer to attackers exploiting transmitter knowledge for semantic encoding and decoding, harvesting the same semantics as the transmitter. Specifically, the attacker who initiates the transmitter knowledge attack has abilities with semantic encoding and decoding, channel encoding and decoding. the attacker exploits transmitter knowledge to train a brand new semantic model, the interpretation of source messages by the attacker will be consistent with the transmitter. Consequently, transmitter knowledge is almost meaningless for semantic model. Furthermore, the transmitter knowledge may have overlapping with receiver, attacker can analyze which part is the receiver knowledge based on the source message.

\subsubsection{Personal Knowledge Assisted Semantics Eavesdropping}
In this case, attackers acquire both personal and global knowledge as well as a matched semantic coding model. It is evident that attackers in this case are highly capable of eavesdropping on both public and private semantics, thereby posing significant challenges on user privacy in SC. Since attackers may hold personal knowledge of transmitters and receivers, the privacy of both sides might be leaked. For transmitters, attackers exploit personal knowledge to infer sensitive information and global knowledge to infer public information. Moreover, the attacker may fully decode the meaning of source messages from the perspective of receivers, particularly for sensitive information of receivers. In addition, attackers can exploit these personal knowledge discrepancies between receivers and transmitters to potentially predict future messages exchanged between the pair. 

By analyzing the above three different kinds of eavesdropping attacks, it is clear that the attacker's understanding of personal knowledge directly determines how much privacy can be stolen. Furthermore, general semantic coding models are usually loosely coupled with knowledge and may not protect privacy well. Therefore, it is necessary to design a new SC framework that includes knowledge management and data encoding/decoding to wisely exploit knowledge to accurately exchange semantics while protecting data privacy. 

\begin{figure*}[ht] % htbp表示将图片放置于当前位置、上方、下方或两侧（如果有足够空间）
    \centering
    \includegraphics[width=\textwidth]{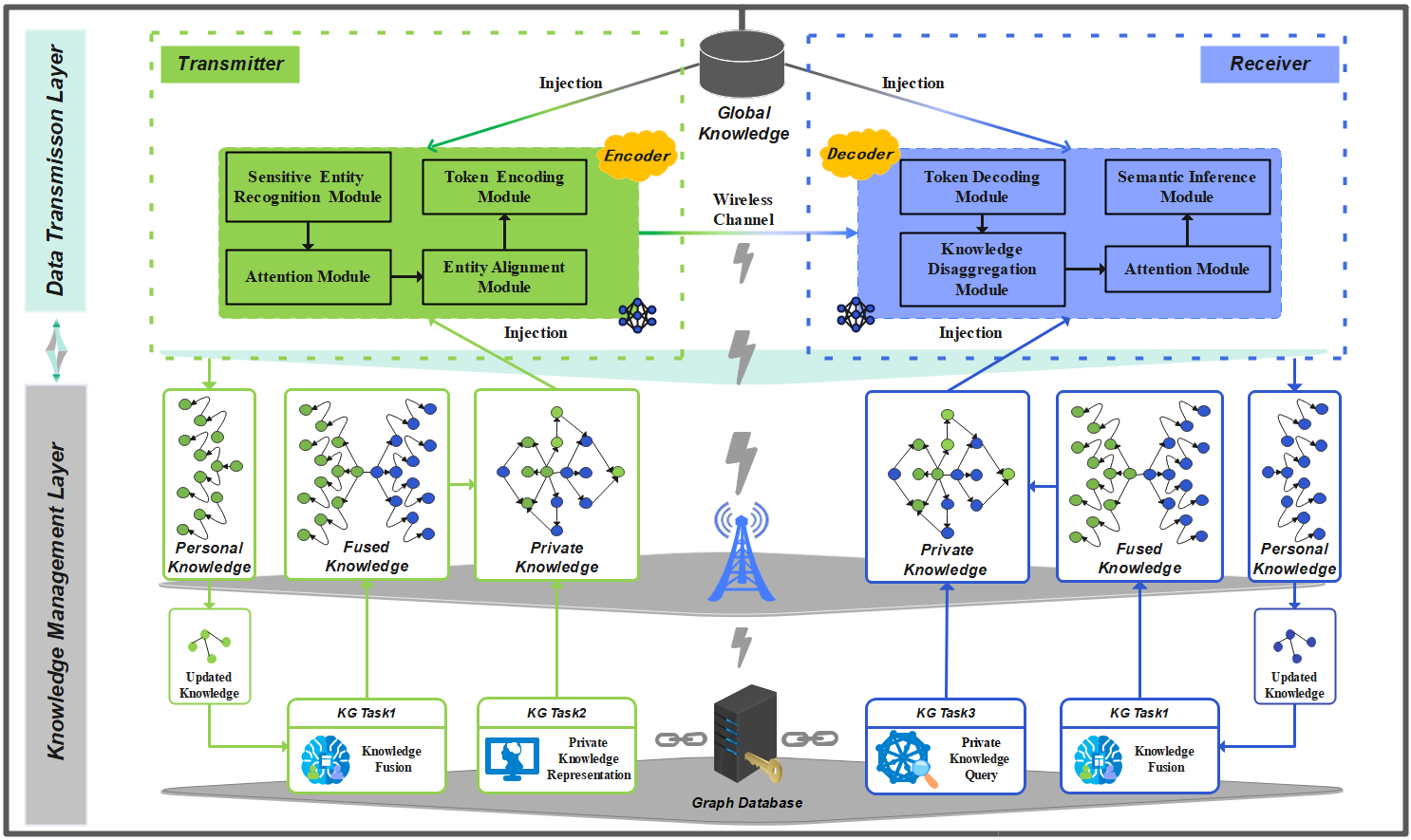} % your_image.jpg为要插入的图片路径
    \caption{Our proposed knowledge-assisted privacy preserving SC framework.} % 添加标题
    \label{fig:33} % 设定引用标签
\end{figure*}

\begin{figure*}[ht] % htbp表示将图片放置于当前位置、上方、下方或两侧（如果有足够空间）
    \centering
    \includegraphics[width=\textwidth]{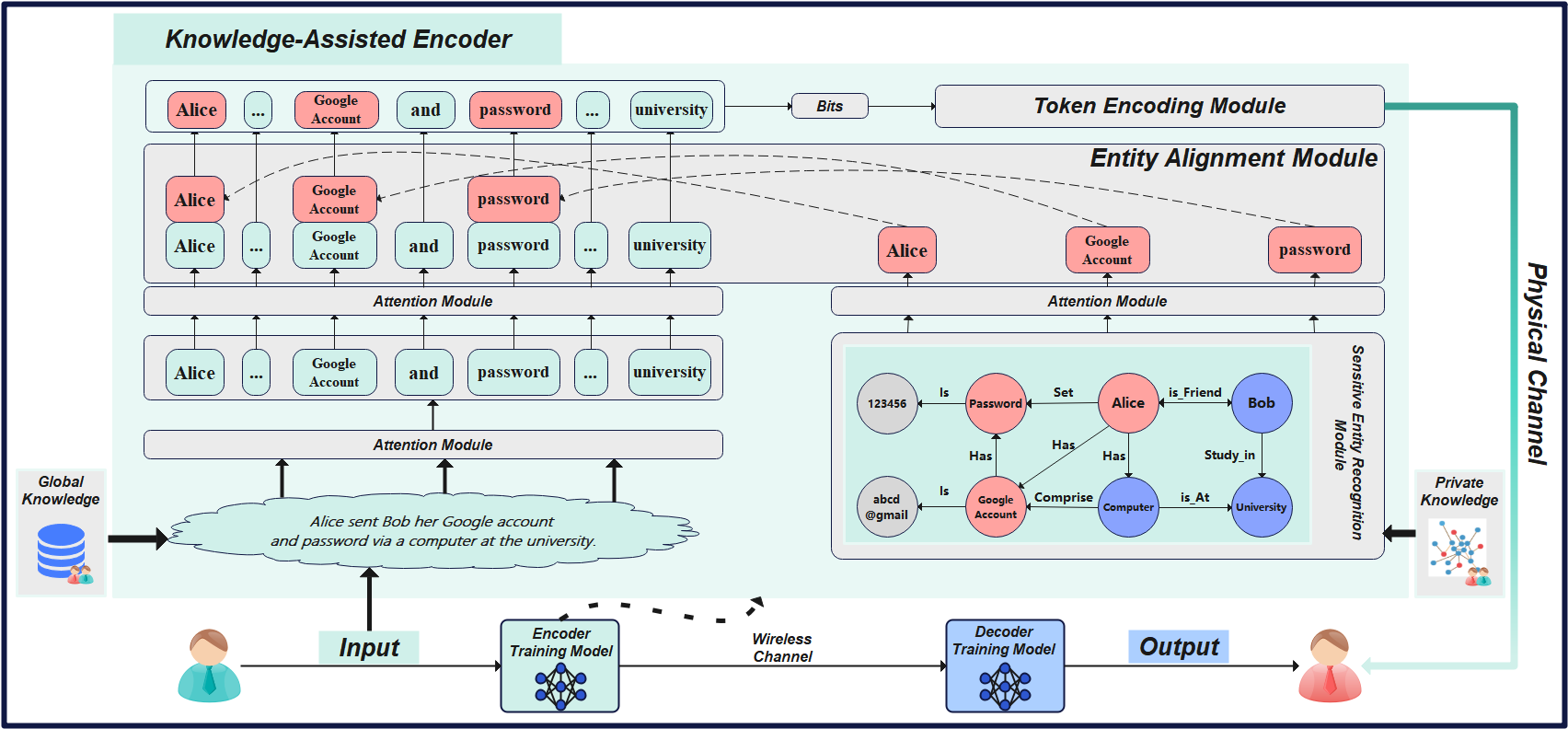}  % your_image.jpg为要插入的图片路径
    \caption{The intrinsic structure of knowledge-assisted encoder networks.} % 添加标题
    \label{fig:2} % 设定引用标签
\end{figure*}
\section{Knowledge-assisted Privacy Preserving SC Framework}
 As shown in Fig. 2, the proposed SC framework consists of two layers: a knowledge management layer and a data transmission layer. The knowledge management layer undertakes knowledge coordination between transmitters and receivers, while the data transmission layer utilizes novel SC transceivers with the assistance of injected knowledge. To illustrate the framework clearly, we will focus on the scenario with one transmission pair in the following.
\subsection{Data Transmission Layer}
Let us consider a transmitter, a receiver, and a trusted wireless network for bit transmission. The wireless network can be untrusted in practice, however, this is out of the scope of our work, and can be solved by some existing wireless security techniques, such as physical layer security.

In this framework, global knowledge is injected into the coding model through word embedding methods, while private knowledge is injected into the coding model through graph embedding methods. Generally, the transmitter extracts the semantic features from source messages by injecting knowledge to minimize the amount of data to be transmitted, and the source message in this way is encoded as a sequence of symbols for transmission via wireless channel. Correspondingly, the receiver decodes the received distorted symbols and recovers the semantics of the original messages with the assistance of its knowledge. Specifically, an encoder located in the transmitter comprises four modules, including sensitive entity recognition, attention, entity alignment, and token encoding module. At the receiver end, a decoder having the same knowledge as that of the encoder is deployed to recover the semantics of source messages from the received channel-distorted bits. This requires another four modules at the receiver side, namely the token decoding module, knowledge disaggregation, attention, and semantic inference. The detailed illustration of the encoder and decoder design will be provided in the next section.

\subsection{Knowledge Management Layer}
As discussed, knowledge plays a unique role in extracting and recovering semantics at both sides. To fully unleash the potential of knowledge, global knowledge is openly accessible as shared datasets, while graph database is utilized as a specialized library for processing personal knowledge and storing private knowledge in the knowledge management layer. Before discussing the process of knowledge management, let us define the concepts of fused knowledge, private knowledge, and updated knowledge, which are specially introduced in our framework.  
\textbf{Fused knowledge} merges the personal knowledge of a transmitter and a receiver through the form of knowledge graph (KG). Here we specify that each entity in this KG represents a basic semantic element, such as place, event, people, etc. An edge connecting two entities in this KG represents the relationships.
\textbf{Private knowledge} is a subgraph of the fused knowledge, including all entities and their relationships with respect to the coming source messages. Moreover, private knowledge may include extra entities related to the source messages due to the structured KG. 
Additionally, \textbf{updated knowledge} is the newly added information in personal knowledge, which ensures the timeliness of personal knowledge. In this way, both fused knowledge and private knowledge can be dynamically updated. 
%In this section, we construct a trustworthy KB around privacy preserving. In Fig. 4, starting with the four basic functions supported by the KG, how to make KG in SC trusted is discussed in technical details, with the intention of obtaining comparable performance to plaintext inputs using privacy knowledge in KBSC while preserving data privacy.

\subsubsection{Knowledge Fusion} This task illustrates the process of merging personal knowledge between the transmitter and the receiver in trusted graph databases, ensuring the same understanding of both sides in SC. To construct this fused knowledge in KG format, knowledge alignment, and knowledge disambiguation are required. 

Knowledge alignment is the process of matching entities with the same name between the transmitter's and receiver's personal knowledge. For example, an entity ``apple" may exist in the personal knowledge of both sides, and we need to replace the two dispersed entities with an entity in the fused KG, thus eliminating redundant knowledge and ensuring its consistency. Private set intersection (PSI) \cite{r20} accomplishes this through hash-encrypted knowledge exchanges, without revealing non-intersecting parts of personal knowledge. 

Knowledge disambiguation is the process of distinguishing knowledge with the same names but different semantics, thus reducing semantic ambiguity in SC. Contextual analysis methods, such as word sense disambiguation \cite{r19} are typically exploited. By matching target entities and their relationships in the context of the source message, the semantics of the target entity can be determined. For instance, in the sentence ``Apple is a company", analyzing the relationship between the entities ``apple" and ``company", ``apple" refers to a brand rather than a fruit. Consequently, fused knowledge is established in a KG format, and the issues of knowledge redundancy and disambiguation can be resolved. Considering the eavesdropping risk of updated knowledge, it is difficult for the attacker to generate the same fused knowledge as the transmitter and receiver due to the lack of personal knowledge.

\subsubsection{Private Knowledge Construction} With the fused knowledge, it ensures the same understanding capability of the transmitter and receiver. However, if we directly inject the fused knowledge into the encoder and decoder, this might be a large-scale and extremely complicated semantic coding model. Moreover, considering that the distribution of a given source message in the fused knowledge may be a subgraph with a very small scope, it is necessary to further clarify the scope of private knowledge in the fused knowledge. This process of private knowledge construction is divided into three steps, namely source message analyzing, source message querying, and minimal KG construction. Source message analyzing is to convert source messages into the format for entity querying, which can be directly realized by natural language processing techniques. Source message querying is to find all entities and relationships related to the source messages in fused knowledge, with the aim of specifying the scope of the private knowledge, graph traversal languages adapted to fused knowledge can effortlessly implement it. Based on this, The entities of source messages in fused knowledge are collected by random walk\cite{r18}, and the minimal spanning tree is utilized to extract the minimal connected subgraph in fused knowledge, removing irrelevant nodes and relationships, thereby optimizing the structure of private knowledge.
\subsubsection{Private Knowledge Distribution} Private knowledge distribution entails that the transmitter and the receiver can obtain the same private knowledge in the graph database. General queries can be carried out directly via graph query language for information retrieval, whereas private knowledge query requires identity authentication before querying. The receiver will be empowered with different permissions through various traditional encryption methods, such as digital signatures and shared keys. Overall, graph databases are effective in ensuring the consistency and integrity of private knowledge in an eavesdropping environment. %this is very nature ans straightforward as the updated personal knowledge will definitely affect/change the fused knowledge, then the "private knowledge".

\begin{figure*}[ht] % htbp表示将图片放置于当前位置、上方、下方或两侧（如果有足够空间）
    \centering
    \includegraphics[width=\textwidth]{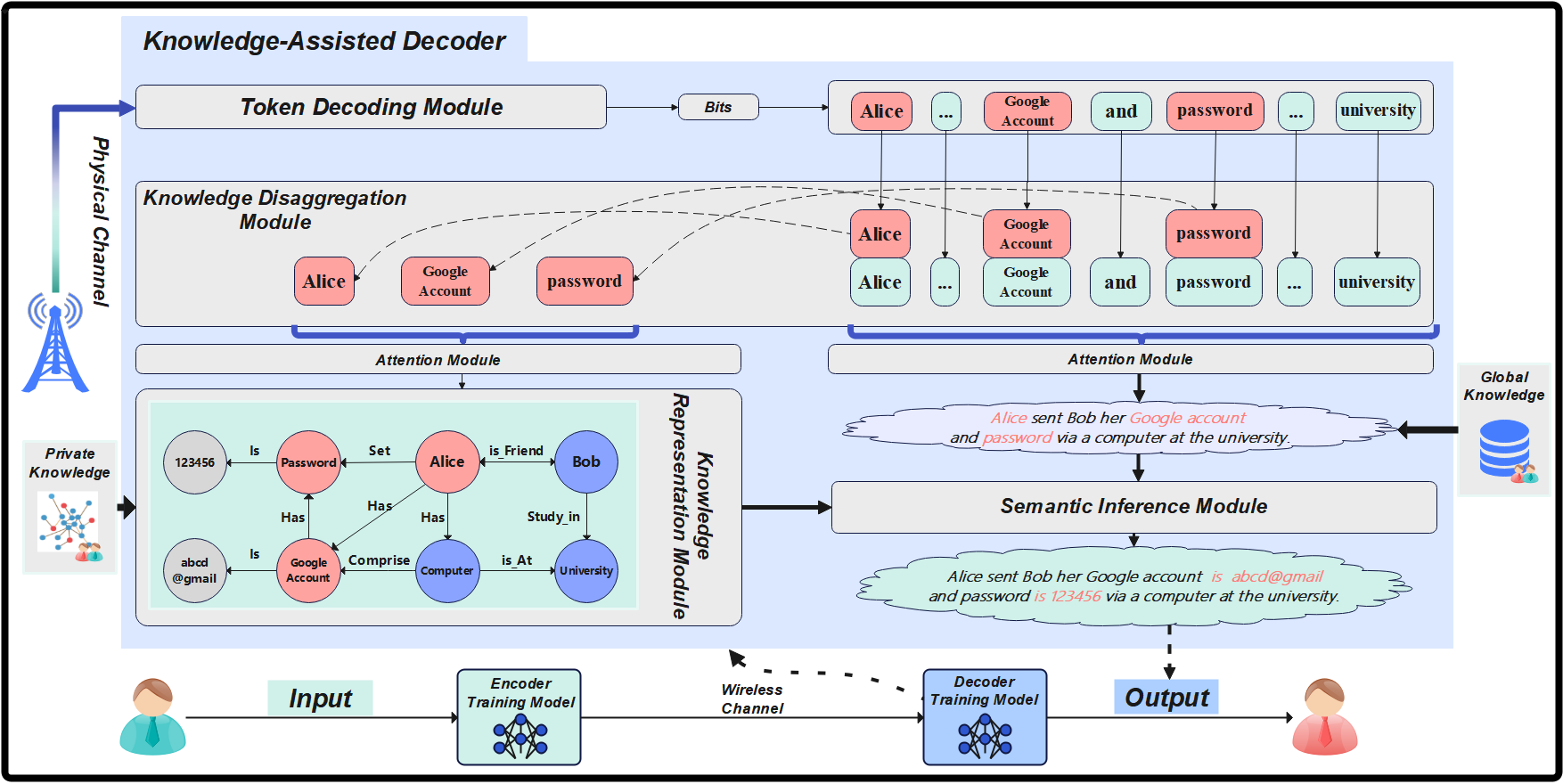} % your_image.jpg为要插入的图片路径
    \caption{The intrinsic structure of knowledge-assisted decoder networks.} % 添加标题
    \label{fig:4} % 设定引用标签
\end{figure*}

\section{Transceiver Design}
In this section, we discuss how to design an SC transceiver with the assistance of knowledge to protect sensitive information.

\subsection{Knowledge-Assisted Encoder Network}
The basic idea in this encoder network design is to leverage global knowledge to compress source messages, while utilizing private knowledge of both sides to protect sensitive information. To this end, the knowledge-assisted encoder shown in Fig. \ref{fig:2} proposed, which is comprised of four modules, namely sensitive entity recognition module, attention module, entity alignment module, and channel coding module. 

 The sensitive entities recognition module consists of two functions. The first function is to screen the potentially sensitive information in private knowledge, Named Entity Recognition (NER) method \cite{r5} is a core task in natural language processing with the goal of recognizing specific entities in private knowledge. NER determines sensitive content in a source message by matching sensitive keywords, such as personal name, Email, ID, etc. This has the advantage of being able to differentiate the intrinsic sensitivity for entities of the same category. For example, "Bob" and "Alice" are both names and should be identified as sensitive information. However, "Alice" has a direct relationship with "Google account" and "password" in private knowledge, thus only "Alice" is identified as sensitive information. The second function is to analyze the relationships among these sensitive entities in private knowledge. Graph neural network (GNN), specialized for processing graph-structured data, could be exploited for identifying internal relationships of sensitive entities in private knowledge by edge classification \cite{r9}. Generally, the complexity of GNN is higher than that of traditional deep neural networks. The computation of GNN involves the sparsity of graph structure data and the aggregation and propagation of neighbor nodes.

The attention module consisting of two layers transforms global knowledge and private knowledge into sequence tokenization. The lower layer attention module captures the basic word-level and syntactic information of the source message, which can be accomplished by natural language processing models, such as BERT. The upper attention module is responsible for integrating tokens between private knowledge and global knowledge, thus enabling to representation of the heterogeneous information of global knowledge and private knowledge in a unified feature space.

The entity alignment module contains original tokens represented by global knowledge, and sensitive tokens represented by private knowledge. The core idea of this module is to find the sensitive tokens, overlaying sensitive tokens over original tokens with the same semantics. Intuitively, the semantic similarity between original tokens and sensitive tokens can be computed directly using traditional methods, such as structural similarity\cite{r8}. Since original tokens are non-structured data, the structural similarity method may cause them to lose their structural properties.  Additionally, its low computational complexity makes it applicable in scenarios with limited wireless resources. Another reliable method is to train a joint embedding model\cite{r6} that puts original tokens and sensitive tokens in the same vector space and evaluates their similarity by calculating the cosine angle of the two vectors, overlaying sensitive tokens into original tokens with similar semantics. However, this method has a high complexity in the training phase and consumes a lot of computing resources and time in scenarios that require large-scale data processing. Finally, a set of tokens jointly represented by private and global knowledge is converted into binary bits and entered into the token encoding module for transmission, ensuring the transmission of sensitive information via different channels precisely. 

The top part of Fig. \ref{fig:5} visualizes the data flow as it passes through the proposed SC encoder. The source message is unstructured data. Text preprocessing decomposes the source message into words or phrases, and transforms the source message into a vector representation through word embedding, extracting the features in the source message. The word or phrase becomes the candidate entity in turn, and the entity that is most relevant to the knowledge is matched according to its feature ranking. Moreover, entity naming identifies sensitive and non-sensitive information in a source message by labeling it with an entity type. A vocabulary is constructed through the relationships between entities, and a set of relationships will be mapped to a unique ID which can be represented by a token. Additionally, which token represents sensitive information can be accomplished by entity replacement. Finally, these tokens are converted into binary bits via token encoding module, and these bits are then transmitted over a noisy wireless channel. 

\subsection{Knowledge-Assisted Decoder Network}
As illustrated in Fig. 4, the token decoding module within the knowledge assisted-decoder is responsible for recovering the original tokens. Subsequently, these tokens are entered within the knowledge disaggregation module. Original tokens and sensitive tokens are recovered via a neural network, which specially obtains the relationships between sensitive and public information. The cooperative interaction between the attention module and the knowledge representation module enables the parsing of sequences containing sensitive information into sensitive tokens, these tokens are then converted back to corresponding entities to be located in private knowledge. These located entities further enable sensitive information to be better inferred at the receiver side. Next, the tokens represented by global knowledge will be recovered directly to source semantics. 

The semantic inference module as the main component of the Knowledge-Assisted decoder refers to the retrieval of sensitive information for source messages in private knowledge, thereby understanding the deep sensitive information. Fine-grained sensitive information helps to locate the initial entity in private knowledge, improving the accuracy of inference. Thus, Rule-based local inference methods \cite{r13} can analyze the edges of sensitive entities to explore new entities and obtain new relationships, inferring new semantics, such as ``Password is 123456" and ``Google Account is abcd@gmail". However, this method may have a relatively long reasoning time and limited accuracy. To deal with these issues, neural networks can be used to model triples and multipaths thus establishing the relationship between sensitive entities and other related entities. By converting the rules into vector operations and applying the neural network method with strong learning ability, the end-to-end differentiable model is realized to ensure inference and generalization. Besides, since harsh channel conditions may cause some core semantics to be lost during the decoding process, the semantic inference module can use private knowledge to infer the missing semantic information, thereby compensating for the information loss caused by channel impairments and noise, and assisting the system to infer reasonable semantics in scenarios with low signal-to-noise ratio.

\begin{figure}[ht] % htbp表示将图片放置于当前位置、上方、下方或两侧（如果有足够空间）
    \centering
    \includegraphics[scale=0.5]{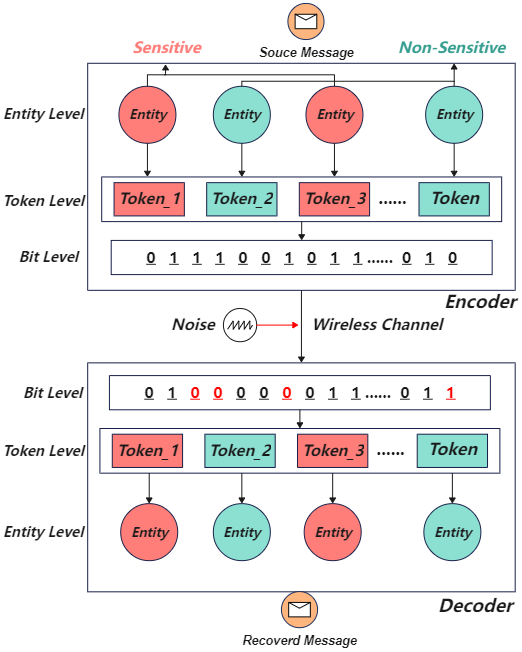}  % your_image.jpg为要插入的图片路径
    \caption{Data flows in the proposed transceiver.} % 添加标题
    \label{fig:5} % 设定引用标签
\end{figure}

As shown in the bottom part of Fig. \ref{fig:5}, the data flow at the decoder part is illustrated. The distorted binary bits are first decoded by the receiver from the received wireless signals, these bit is converted to a token by the token decoding module. A vocabulary has been constructed in the encoder during the conversion of entities to tokens. Therefore, the same applies to decoders, and the token will be converted to the corresponding entity. Finally, semantic information can be obtained by inferring various relationships of entities.

\subsection{Analysis on Eavesdropping Attacks}

Despite we assume a trusted graph database in knowledge management of our framework, both global and personal knowledge may still be stolen and exploited by eavesdroppers. As a result, all three kinds of attacks discussed in Section II.B could occur. Let us demonstrate how our proposed SC framework can protect from the three eavesdropping attacks.

In the case of bit eavesdropping, attackers with no matched semantic coding model can only intercept bits via the wireless channel, and decode these bits at a syntactic level. Despite the attackers can detect the information exchanged between the two ends, they cannot obtain semantic information. Therefore, limited privacy could be leaked.

As discussed, the second type of global knowledge assisted semantics eavesdropping attack assumes that the attackers are equipped with a capable decoder along with the global knowledge. This attack may manifest in the bit and token levels. Second, we match global knowledge with private knowledge in the entity alignment module for heterogeneous information, tokens cannot be turned into source messages without private knowledge. The decoding model from attackers may not have this function, much less private knowledge, resulting in a large number of errors in the decoding process. Thus, it is almost impossible for the attacker to obtain valuable public information from global knowledge. Our designed knowledge management layer and knowledge-assisted encoder provide dual protection against global knowledge assisted semantics eavesdropping.

Personal knowledge assisted semantics eavesdropping attacks empower attackers to decode public and private information. This attack has a great impact on the bit level and token level, and poses a challenge to the entity level. Similar to global knowledge assisted eavesdropping attacks, the graph database stores personal knowledge of the transmitter and receiver, radically reducing the risk of eavesdropping at the bit level. The difference is that we fuse the personal knowledge into private knowledge at the knowledge management layer, and attackers cannot get the private semantics even if it decodes through the personal knowledge, avoiding token level threats. However, smart attackers are likely to fuse personal knowledge to obtain private knowledge, launching a challenge to the entity level. To this end, we design a sensitive entity recognition module to further screen sensitive entities in private knowledge, distinguishing the definition of sensitive information in the same private knowledge between eavesdroppers and receivers effectively, which ensures the transmission of private semantics precisely.

\section{Open Issues and Outlook}
Despite the expected prospects of the proposed knowledge-assisted privacy preserving SC framework, there are still some challenges that need to be addressed before relishing its full potential.

\textbf{Dynamic Channel Condition:}
The transceivers in our proposed SC framework are assumed to be trained under a given channel condition. However, wireless channel conditions in practical scenarios are typically dynamic, and this may significantly degrade both data transmission accuracy and privacy levels. For example, severe bit error rate may lead to incorrect tokens decoded at receivers, while excellent channel conditions might enable eavesdroppers to decode excessive public information, potentially compromising privacy levels. Therefore, enhancing the generalization capability of transceivers in the proposed SC framework should be another non-trivial challenge. Some prompt-based learning algorithms can be exploited to timely adjust or fine-tune parameters in the coding model to adapt channel dynamics. 

\textbf{Scalability of the knowledge management layer:} With the increase of network nodes and devices, the graph database needs to process more data, especially in the process of frequent knowledge updates. In the case of unstable networks or frequent node changes, it is particularly important to synchronize knowledge updates between different nodes and maintain the consistency of knowledge in the transceiver~\cite{10271127}. In addition, KG tasks are compute-intensive, and these tasks have different bandwidth requirements. In low-latency scenarios, how to precisely distribute these tasks to different nodes and ensure balanced computing load is an issue.

\textbf{Overwhelmed Private Knowledge:} 
Compared to global knowledge, the information volumes and training parameters in private knowledge are relatively insignificant, thus private knowledge may be overwhelmed by global knowledge. In this case, the information represented by sensitive entities in private knowledge will be replaced by global knowledge, when performing entity alignment module. A possible approach is to introduce complex attention mechanisms into the model to better utilize the knowledge. Additionally, considering varying knowledge preferences and limited knowledge storage capacities among different SC users, how to realize efficient knowledge construction to maximize the overall semantic performance in SC-based networks becomes rather challenging.

\textbf{High Computational Burden on Mobile Devices:}
Private knowledge may contain numerous entities and edges, and embedding these into a coding model requires dealing with complex graph structures. This significantly increases the computational complexity of the model. Therefore, the training time could be prolonged, affecting the iteration and convergence speed of the model. To alleviate the computing burden on mobile devices, incremental learning and distributed computing can be utilized \cite{r21}.

\textbf{Lack of Standardized Metrics of Privacy in SC:}
Measuring the level of privacy in SC is another challenge. Traditional metrics from information theory, such as entropy and mutual information, may not be directly applied to measure privacy in SC. Entropy cannot distinguish between sensitive and non-sensitive information, and if a high entropy of a dataset, this does not mean a high privacy risk. Similarly, mutual information can measure the amount of information between two variables/messages, but it cannot specifically measure which information poses a threat to privacy. To define a suitable metric, exploiting differential privacy could be a potential way.

%\textbf{Privacy-Preserving Channel Selection:} Precise channel evaluation can avoid privacy snooping to the greatest extent possible, establish a channel evaluation model, score the available channels based on privacy protection strength, communication efficiency and other factors, monitor the channel status in real time, and dynamically adjust the channel policy, a small communication cost in exchange for privacy and security is acceptable.

\section{Conclusion}
In this article, we explored the utilization of knowledge to enhance data privacy in SC networks. We began by analyzing the potential three kinds of eavesdropping attacks in SC, then proposed our privacy-preserving SC framework, consisting of data transmission layer and knowledge management layer. Especially, we elaborated the transceiver design in this framework to ensure an accurate data transmission at a semantic level while protecting sensitive information from eavesdroppers by exploiting private knowledge at the transmitter and receiver sides. This work can be seen as a pioneer in exploiting knowledge to enhance data privacy in SC.

        \bibliographystyle{IEEEtran}
	\bibliography{ref}
\vspace{-10pt}
\end{document}